\documentclass[preprint,showpacs,preprintnumbers,amsmath,amssymb, superscriptaddress]{revtex4}

\usepackage{textcomp}
\usepackage{makeidx}
\usepackage{amsmath}
\usepackage{subfigure}
\usepackage{amssymb}
\usepackage{hyperref}
\usepackage{cleveref}
\usepackage{graphicx}
\usepackage[utf8]{inputenc}
\usepackage{float}
\usepackage{color}
\usepackage{epsfig}
\usepackage{eucal}
\usepackage{amsmath}
\usepackage{tabularx}
\usepackage{tabulary}
\usepackage{graphics}
\usepackage{blindtext}
\usepackage{dblfloatfix}
\begin{document}

\title{Anisotropic fluid spheres satisfying Karmarkar condition}

\author{Francisco Tello-Ortiz}
\email{francisco.tello@ua.cl}
\affiliation{Departamento de F\'isica, Facultad de ciencias básicas, Universidad de Antofagasta, Casilla 170, Antofagasta, Chile.}


\begin{abstract}
In this work we obtain an analytic and well behaved solution to Einstein's field equations describing anisotropic matter distribution. It's achieved in the embedding class one spacetime framework using Karmarkar's condition. We ansatz the metric potential $g_{tt}$ corresponding to Wyman IIa solution. The obtained model is representing some strange star candidates such as Cen X-3, LMC X-4 and 4U 1538-52. A graphical analysis shows that our model is free from physical and geometric singularities and satisfies all the physical admissibility conditions from the physical and mathematical point of view. 
\end{abstract}

\maketitle

\section{Introduction}

Due to the highly non linear behavior of Einstein's field equations, it remains a great challenge to obtain solutions that meet the requirements in order to be an admissible solutions from both physical and mathematical point of view \cite{delgaty}. The early works by Schwarzschild \cite{schwar}, Tolman \cite{tolman} and Oppenheimer and Volkoff \cite{oppenheimer} on self gravitating isotropic fluid spheres, established (as was pointed out by Murad \cite{murad}) two \emph{classical approaches} that can be followed in order to solve Einstein's equations. The first one consists in make a suitable assumption for one of the metric functions or for the energy density. This leads to the remaining unknown variables, that is the isotropic pressure and the other metric potential. However, in the framework of this scheme not always is possible to obtain an acceptable solution (sometimes one obtain nonphysical pressure-density configurations). The second approach start with an equation of state which is integrated (iteratively from the center of the compact object) until the pressure vanished indicating that the object surface has been reached. As before, this scheme also presents some drawbacks since it does not always lead to a closed form of the solutions.

On the other hand, a stellar configuration not necessary needs to meet the isotropic condition at all (equal radial $p_{r}$ and tangential $p_{t}$ pressure). In fact the theoretical studies by Ruderman \cite{ruderman}, Canuto \cite{canuto,canuto1,canuto2} and Canuto et. al \cite{canuto3,canuto4,canuto5,canuto6}  revealed that when the matter density is much higher than the nuclear density, it may be anisotropic in nature and must be treated relativistically. So, relaxing the isotropic condition and allow the presence of anisotropies (it leads to unequal radial and tangential pressure $p_{r}\neq p_{t}$) within the stellar configuration represent a more realistic situation in the astrophysical sense. Furthermore, the pioneering work by Bowers and Liang about  \cite{bowers} local anisotropic properties for static spherically symmetric and relativistic configurations, gave rise an extensive studies within this framework, specifically studies focused on the dynamical incidence of the anisotropies in the arena of equilibrium and stability on collapsed structures \cite{heintz,cosenza,cosenza1,bayin,krori,herre,jponce,jponce1,chan3,hbondi,chan1,herrera,chan2,mehra,Herrera,herre1,herre2,kdev,kdev1,kdev2,ivanov,harko,harko1,harko2,abreu,via,ivanov1}. Moreover, as Mak and Harko have argued \cite{harko1}, anisotropy can arise in different contexts such as: the existence of a solid
core or by the presence of type 3A superfluid \cite{kippen}, pion condensation \cite{sawyer} or different kinds of phase transitions \cite{sokolov}. 

Over the years many researches have successfully addressed the study and understanding of the role played by the anisotropy in stellar interiors \cite{schunk,ray,usov,mello,negreiros,varela,rahaman1,rahaman2,rahaman3,rahaman,hassan1,maurya7,bhar1,rahaman4,kileba1,hassan,shee,monadi,smaurya1,smaurya2,deb1,Ovalle8,Ovalle,role,jasim,Tello1,Gabbanelli,camilo,sharif,Ovalle9,Tello,Tello2,deb,kileba}. The presence of anisotropy introduces several features in the matter distribution, e.g. if we have a positive anisotropy factor $\Delta\equiv p_{t}-p_{r}>0$, the stellar system experiences a repulsive force (attractive in the case of negative anisotropy factor) that counteracts the gravitational gradient. Hence it allows the construction of more compact objects
when using anisotropic fluid than when using isotropic fluid \cite{mehra,ivanov,harko1}. Furthermore, a positive anisotropy factor enhances the stability and equilibrium of the system.

In recent years, the use of the embedding class one spacetime via Karmarkar \cite{kar1} condition as a systematic and powerful method to obtain new and relevant solutions from the Einstein field equations, has increased \cite{kar2, kar3, kar4, kar5, kar6,kar7,kar8,kar9,kar10,kar11,kar12,kar13,kar14,kar15,kar16,kar17,kar18,kar19,kar20,kar21,kar22,kar23,kar24,kar25,kar26,kar27,kar28,kar29,kar30,kar31,kar32,kar33,kar34}. For a spherically symmetric $4$-dimensional spacetime, the Karmarkar condition in
terms of curvature components takes the form
\begin{equation}\label{karmar}
R_{1010}R_{2323}-R_{1212}R_{3030}=R_{1220}R_{1330}.  
\end{equation}
However, as was pointed out by Sharma and Pandey \cite{pandey}, the above condition is not enough to spherically symmetric $4$-dimensional spacetime render to be class one. So, in order to be class one a $4$-dimensional manifold must satisfies 
\begin{equation}\label{pandshar}
R_{2323}\neq 0,    
\end{equation}
along with (\ref{karmar}). As we will see later equations (\ref{karmar}) and (\ref{pandshar}) arrive to a particular differential equation that links the
two metric components $e^{\nu}$ and $e^{\lambda}$. Therefore, one only needs to
specify one of the metric potential. Finally the rest of the physical quantities like pressure, density, sound speed,
anisotropy, etc. can be completely determined from $e^{\nu}$ and $e^{\lambda}$.

The outline of the paper is as follows: Section \ref{section2} presents the Einstein's field equations for anisotropic matter distributions and the approach followed in order to solve them, in section \ref{section3} we match the obtained model with the exterior spacetime given by Schwarzschild metric, in order to obtain the constant paremeters. Section \ref{paf} provides the basic requeriments that a well behaved solution must fulfills, in section \ref{section5} we analyze the requirements listed in the previous section such as the positiveness of the metric potentials and the thermodynamic observables within the star. Section \ref{section6} is devoted to the study of the equilibrium via TOV equation and the stability of the present model.  Finally section \ref{section7} discusses and concludes the work.

\section{Einstein's field equations}\label{section2}
In Schwarzschild like coordinates the interior of a compact static spherically symmetric object is described by the following line element
\begin{equation}\label{metric}
ds^{2}=e^{\nu}dt^{2}-e^{\lambda}dr^{2}-r^{2}\left(d\theta+\sin^{2}\theta d\phi\right),    
\end{equation}
where $\nu=\nu(r)$ and $\lambda=\lambda(r)$ are purely radial functions only. Assuming an anisotropic matter distribution within the stellar configuration, the corresponding energy-momentum tensor is 
\begin{equation}\label{matter}
T^{\nu}_{\mu}=\left(\rho+p_{t}\right)U^{\nu}U_{\mu}-p_{t}\delta^{\nu}_{\mu}-\left(p_{t}+p_{r}\right)V^{\nu}V_{\mu},    
\end{equation}
being $U^{\nu}$ the four-velocity $e^{\nu(r)/2} U^{\theta} = \delta^{\theta}_{0}$, while $V^{\theta}$ is a unit spacelike vector in the radial direction
$V^{\theta}= e^{-\lambda/2}\delta^{\theta}_{1}$, which is orthogonal to $U^{\theta}$. Here $\rho$ is the matter density,
$p_{r}$ is the radial and $p_{t}$ is transverse pressure of the fluid in the orthogonal direction to $p_{r}$.
Then, the Einstein field equations for the line element (\ref{metric})
and the energy-momentum tensor (\ref{matter}) are given by
\begin{eqnarray}\label{effectivedensity}
8\pi {\rho}&=&\frac{1}{r^2}-e^{-\lambda}\left(\frac{1}{r^2}-\frac{\lambda^{\prime}}{r}\right),\\\label{effectiveradialpressure}
8\pi {p}_{r}&=&-\frac{1}{r^2}+e^{-\lambda}\left(\frac{1}{r^2}-\frac{\nu^{\prime}}{r}\right),\\\label{effectivetangentialpressure}
8\pi {p}_{t}&=&\frac{1}{4}e^{-\lambda}\left(2\nu^{\prime\prime}+\nu^{\prime2}-\lambda^{\prime}\nu^{\prime}+2\frac{\nu^{\prime}-\lambda^{\prime}}{r}\right).
\end{eqnarray}
The primes denote differentiation with respect to the radial coordinate $r$. From now on relativistic
geometrized units are employed, that is $c=G=1$. 

So, combining the expressions (\ref{effectiveradialpressure}) and (\ref{effectivetangentialpressure}) we get
\begin{equation}\label{anisotropy}
\begin{split}\Delta\equiv {p}_
{t}-{p}_{r}=  
e^{-\lambda}\bigg[\frac{\nu^{\prime\prime}}{2}-\frac{\lambda^{\prime}\nu^{\prime}}{4}+\frac{\nu^{\prime2}}{4}-\frac{\nu^{\prime}+\lambda^{\prime}}{2r}+\frac{e^{\lambda}-1}{r^{2}}\bigg],
\end{split}
\end{equation}
where $\Delta$ is called the anisotropy factor which measures the anisotropy inside the star.
\subsection{Karmarkar condition}

At this stage we have five unknown function, namely $\nu$, $\lambda$, $\rho$, $p_{r}$ and $p_{t}$. In order to solve the system of equations (\ref{effectivedensity})-(\ref{effectivetangentialpressure}) we employ the method used by Karmarkar \cite{kar1} where the obtained solutions are classified as class one spacetime. In this
method the Riemann curvature tensor $R_{\alpha\beta\mu\nu}$ satisfies a particular equation that
finally links the two metric component $e^{\nu}$ and $e^{\lambda}$ in a single equation, i.e. the
two metric components are dependent on each other.

The non zero component of the Riemann curvature tensor for the line element (\ref{metric}) are

\begin{eqnarray}\label{1}
R_{1010}&=&-e^{\nu}\left(\frac{\nu^{\prime\prime}}{2}-\frac{\lambda^{\prime}\nu^{\prime}}{4}+\frac{\nu^{\prime2}}{4}\right), \\
R_{2323}&=&-e^{-\lambda}r^{2}\sin^{2}\theta\left(e^{\lambda}-1\right),\\
R_{3030}&=&-\frac{r}{2}\nu^{\prime}e^{\nu-\lambda}\sin^{2}\theta, \\ \label{2}
R_{1212}&=&\frac{r}{2}\lambda^{\prime},
\end{eqnarray}
then all the above components of Riemann curvature satisfy Karmarkar condition \cite{kar1}
\begin{equation}\label{karmarkar}
R_{1010}R_{2323}-R_{1212}R_{3030}=R_{1220}R_{1330}.  
\end{equation}
However, as was pointed out by Pandey and Sharma \cite{pandey}, the above condition is only a necessary one but it is not sufficient to spacetime becomes class one. In order to be class one a spacetime must satisfies (\ref{karmarkar}) along with $R_{2323}\neq0$ \cite{pandey}.

On substituting (\ref{1})-(\ref{2}) in (\ref{karmarkar}) we obtain the following differential equation
\begin{equation}\label{differenctial}
2\frac{\nu^{\prime\prime}}{\nu^{\prime}}+\nu^{\prime}=\frac{\lambda^{\prime}e^{\lambda}}{e^{\lambda}-1},    
\end{equation}
with $e^{\lambda}\neq 1$. Solving (\ref{differenctial}) we arrive to 
\begin{equation}\label{lambda}
e^{\lambda}=1+C^{2}\nu^{\prime2}e^{\nu},  
\end{equation}
where $C$ is an integration constant. Expression (\ref{lambda}) establishes a relationship between the metric potentials $e^{\nu}$ and $e^{\lambda}$.

\subsection{Relativistic embedding class one solution}

To solve the above equation (\ref{lambda}), we have chosen the potential $e^{\nu}$ corresponding to the Wyman IIa solution \cite{wyman} (it corresponds to a special case of Tolman VI solution \cite{tolman})
\begin{equation}\label{nu}
e^{\nu}=(A-Br^{2})^{2},    
\end{equation}
where A and B are constant parameters. 
Employing (\ref{nu}) into (\ref{lambda}) we get
\begin{equation}\label{lamb}
e^{\lambda}=1+16B^{2}C^{2}r^{2}.    
\end{equation}
The obtained potential (\ref{lamb}) resemble us $g_{rr}$ Finch-Skea potential \cite{fs}. So, the class one spacetime reads
\begin{equation}\label{metric}
ds^{2}=\left(A-Br^{2}\right)^{2}dt^{2}-\left(1+16B^{2}C^{2}r^{2}\right)dr^{2}-r^{2}d\Omega^{2},    
\end{equation}
where $d\Omega^{2}\equiv\sin^{2}\theta d\phi^{2}+d\theta^{2}$.
On using (\ref{nu}) and (\ref{lamb}), we can rewrite the expression
of $\rho$, $p_{r}$, $p_{t}$ and $\Delta$ as 
\begin{eqnarray}\label{rho}
\rho(r)&=&\frac{2\left(16B^{2}C^{2}r^{2}+3\right)B^{2}C^{2}}{\pi\left(16B^{2}C^{2}r^{2}+1\right)^{2}} \\ \label{pr}
p_{r}(r)&=&\frac{\left(4B^2C^2r^2-4ABC^2-1\right)B}{2\pi\left(A-Br^2\right)\left(16B^2C^2r^2+1\right)} \\ \label{pt}
p_{t}(r)&=&\frac{\left(4B^2C^2r^2+4ABC^2+1\right)B}{2\pi\left(Br^2-A\right)\left(16B^2C^2r^2+1\right)^{2}}\\
\Delta(r)&=&\frac{4B^3r^2C^2\left(8B^2C^2r^2-8ABC^2-1\right)}{\pi\left(Br^2-A\right)\left(16B^2C^2r^2+1\right)^2}
\end{eqnarray}
\section{Junction conditions}\label{section3}
In order to find the arbitrary constants $A$, $B$ and $C$ we must match our interior solution (\ref{metric}) to
the exterior Schwarzschild solution at the boundary of the star. The line element of the exterior Schwarzschild solution \cite{schwar} is given by
\begin{equation}\label{Schwarzschild}
ds^{2}=\left(1-\frac{2M}{r}\right)dt^{2}-\left(1-\frac{2M}{r}\right)^{-1}dr^{2} 
-r^{2}d\Omega^{2}.  
\end{equation}
 For this purpose we will use the Israel-Darmois junction conditions \cite{Israel,darmois}. Now at the boundary $r=R$ the coefficients of $g_{tt}$ and $g_{rr}$ all are continuous. This implies
\begin{equation}\label{match1}
\left(A-BR^{2}\right)^{2}=1-\frac{2\tilde{M}}{R},
\end{equation}
\begin{equation}\label{match2}
1+16B^{2}C^{2}R^{2}=\left(1-\frac{2\tilde{M}}{R}\right)^{-1}.
\end{equation}
On the other hand, the null radial pressure condition at the boundary  
\begin{equation}\label{secondfundamental}
p_{r}(R)=0,    
\end{equation}
leads to
\begin{equation}\label{B}
C^{2}=\frac{1}{4B\left(A-BR^{2}\right)}.
\end{equation}
Equations (\ref{match1}), (\ref{match2}) and (\ref{B})  are the necessary and sufficient conditions to determine the constants $A$, $B$ and $C$. In addition, the values of the mass $\tilde{M}$ and the radius $R$ have been established based on the obtained data from some strange star candidates such as: Cen X-3, LMC X-4 and 4U 1538-52 \cite{tapa}. In table \ref{table1} are displaying the values of the constant parameters $A$, $B$ and $C$ for each strange star candidate. 
\section{Physical admissibility features}\label{paf}
In order to be physically meaningful, the interior solution for static fluid spheres must satisfy some more general physical requirements.
The following conditions have been generally recognized to be crucial for anisotropic fluid spheres \cite{Herrera}
\begin{enumerate}
 \item The solution should be free from physical and geometric singularities and non zero positive values of $e^{\lambda}$ and $e^{\nu}$ i.e. $(e^{\lambda})_{r=0}=1$ and $e^{\nu}>0$.
 \item The radial pressure $p_{r}$ must be vanishing but the tangential pressure $p_{t}$ may not vanish at the boundary $r=R$ of the sphere. However the radial pressure equal to the tangential pressure at the centre of the fluid sphere.
 \item The density $\rho$ and pressures $p_{r}$, $p_{t}$ should be positive inside the star.
 \item $\left(\frac{dp_{r}}{dr}\right)_{r=0}=0$ and $\left(\frac{d^{2}p_{r}}{dr^{2}}\right)_{r=0}<0$  so that
pressure gradient $\frac{dp_{r}}{dr}$ is negative for $0<r\leq R$.
 \item $\left(\frac{dp_{t}}{dr}\right)_{r=0}=0$ and $\left(\frac{d^{2}p_{t}}{dr^{2}}\right)_{r=0}< 0$ so that pressure
gradient $\frac{dp_{t}}{dr}$ is negative for $0<r\leq R$.
 \item $(\frac{d\rho}{dr})_{r=0}=0$ and $\left(\frac{d^{2}\rho}{dr^{2}}\right)_{r=0}<0$ so that density
gradient $\frac{d\rho}{dr}$ is negative for $0<r\leq R$.
The condition (4), (5) and (6) imply that pressure and density should be maximum at the center and monotonically decreasing towards the surface.
 \item  Inside the static configuration the speed of sound should be less than the speed of light, i.e.
$0\leq\sqrt{\frac{dp_{r}}{d\rho}}<1$ and  $0\leq\sqrt{\frac{dp_{t}}{d\rho}}<1$.
In addition to the above, the velocity of sound should be decreasing towards the surface. i.e.
$\frac{d}{dr}\left(\frac{dp_{r}}{d\rho}\right)<0$
or $\left(\frac{d^{2}p_{r}}{d\rho^{2}}\right)>0$ and 
$\frac{d}{dr}\left(\frac{dp_{t}}{d\rho}\right)<0$ or $\left(\frac{d^{2}p_{t}}{d\rho^{2}}\right)>0$ for
$0\leq r \leq R$ i.e. the velocity of sound is increasing with the increase of density.
\item A physically reasonable energy-momentum tensor
has to obey the null energy condition (NEC), weak energy condition (WEC), strong energy condition (SEC) and the dominant energy condition (DEC). 
\item The central red shift $Z_{0}$ and surface red shift $Z_{R}$ should be positive and finite i.e. $Z_{0}= \left[e^{-\nu(r)/2}-1\right]_{r=0}>0$ and $Z_{R} =\left[e^{\lambda(r)/2} - 1\right]_{r=R}>0$ and both should be bounded.
\end{enumerate}
\section{Physical analysis}\label{section5}
In this section we will analyze the foregoing criteria of the section \ref{paf} in order to investigate if the present model is admissible and a well behaved solution to Einstein's field equations.   
\subsection{Metric potentials and thermodynamic observables}
\begin{figure}[H]
\includegraphics[scale=2.0]{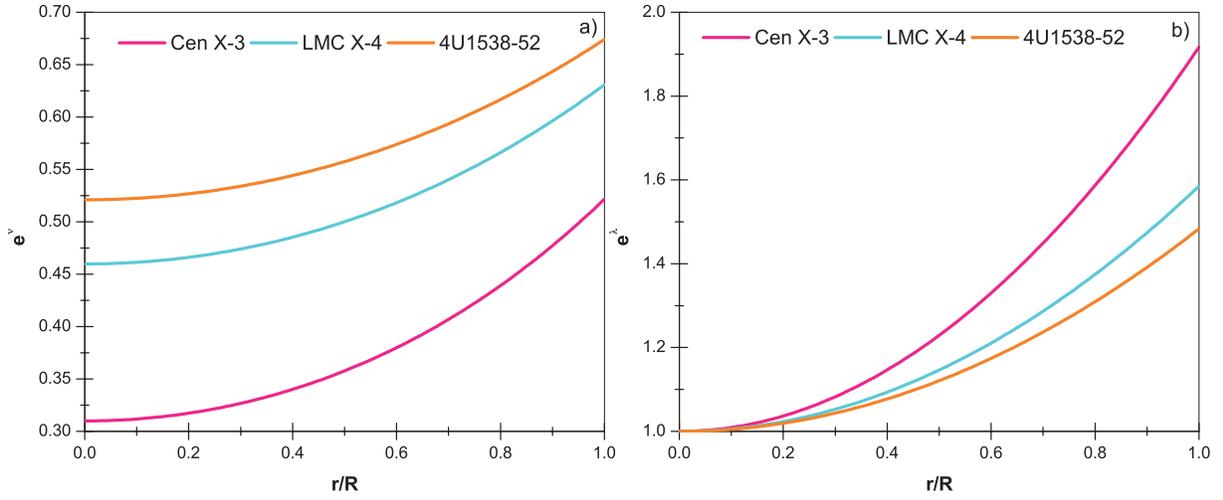}
\caption{Variation of metric potentials with the dimensionless radial coordinate
for Cen X-3, LMC X-4 and 4U1538-52 for
the parameters given in table \ref{table1}.}
\label{pots}
\end{figure}

\begin{figure}[H]
\includegraphics[scale=2.0]{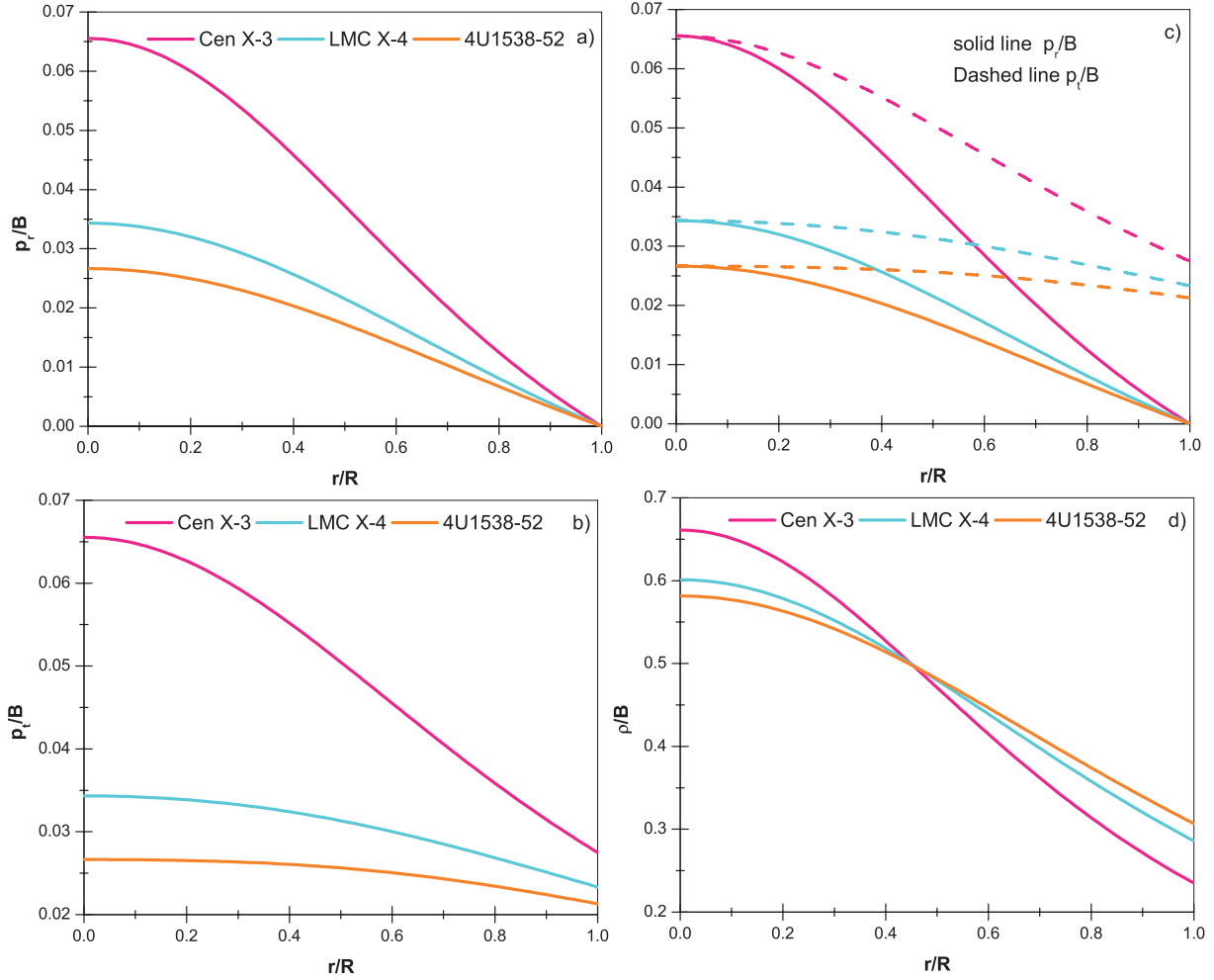}
\caption{Variation of the dimensionless thermodynamic observables with the fractional radial coordinate
for Cen X-3, LMC X-4 and 4U1538-52 for
the parameters given in table \ref{table1}. Panel c) shows how the radial $p_{r}$ and tangential $p_{t}$ pressure drift apart towards the surface.}
\label{thermo}
\end{figure}

\begin{figure}[H]
\includegraphics[scale=2.0]{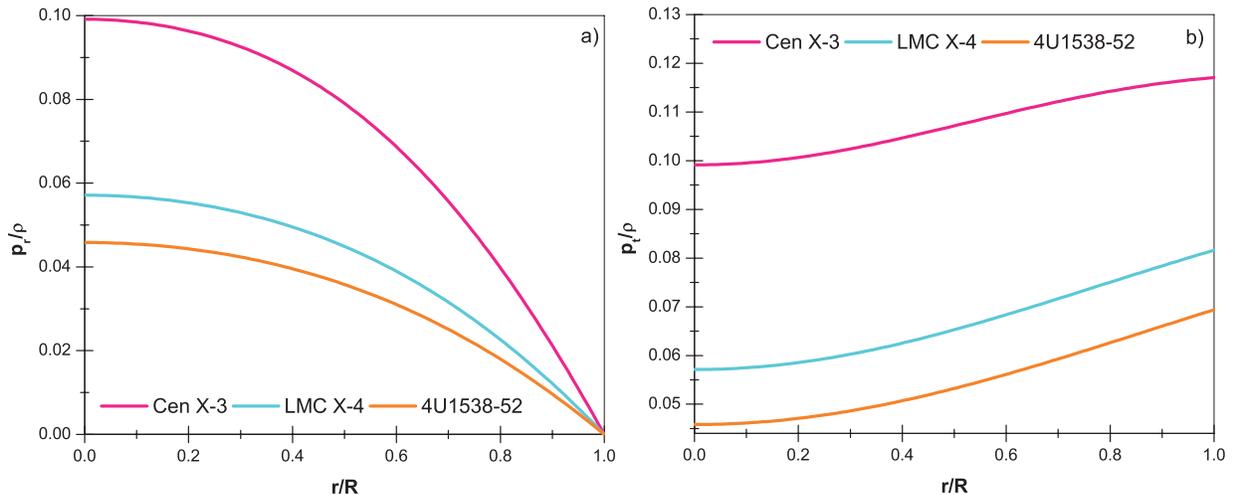}
\caption{Zeldovich's condition against the dimensionless radial coordinate
for Cen X-3, LMC X-4 and 4U1538-52 for
the parameters given in table \ref{table1}.}
\label{zeldovich}
\end{figure}

From expressions (\ref{nu}) and (\ref{lamb}) is clear that the present model is free from physical and geometric singularities as can be seen evaluating at the center $r=0$ of the compact configuration 
\begin{equation}
e^{\lambda(r)}|_{r=0}=1 \quad \mbox{and} \quad e^{\nu(r)}|_{r=0}=A^{2}.    
\end{equation}
Fig. \ref{pots} shows the positive monotonically increasing behaviour of both metric potentials (\ref{nu}) and (\ref{lamb}) within the compact star.
Respect to the density energy $\rho$, radial $p_{r}$ and tangential $p_{t}$ pressures they must have their maximum values at the center of the star and monotonically decreasing behaviour towards the surface. Moreover, the radial pressure $p_{r}$ must vanish at the boundary $\Sigma$, defined by $r=R$. The central values of $\rho$, $p_{r}$ and $p_{t}$ can be obtained from expressions (\ref{rho}), (\ref{pr}) and (\ref{pt}) yielding to 
\begin{eqnarray}\label{centralrho}
\rho(0)&=&\frac{6B^{2}C^{2}}{\pi}>0, \\ \label{centralp}
p_{r}(0)&=&p_{t}(0)=-\frac{B}{2\pi A}\left(4ABC^{2}+1\right)>0.
\end{eqnarray}
From (\ref{centralp}) we get 
\begin{equation}
AB<-\frac{1}{4C^{2}},    
\end{equation}
and using Zeldovich's condition \cite{zeldo} 
\begin{equation}
\frac{p_{r}(0)}{\rho(0)}\leq 1 \Rightarrow AB\geq -\frac{1}{16C^{2}},
\end{equation}
so
\begin{equation}
-\frac{1}{16C^{2}}\leq AB<-\frac{1}{4C^{2}}.    
\end{equation}
Therefore, any A or B should be negative in order to ensure the positiveness of $p_{r}$ and $p_{t}$ inside the configuration. Fig. (\ref{thermo}) shows that all the above quantities are well behaved within the star. At this stage is worth mentioning that the present model exhibits a positive anisotropy factor $\Delta$, it can be seen in figure (\ref{thermo}) panel $c)$ where $p_{t}>p_{r}$ then $\Delta>0$. Thus the object is subject to a repulsive force that counteracts the gravitational gradient, this fact allows the construction of more compact structure \cite{mehra}. Fig. \ref{anisotropy} shows the behaviour of the anisotropy factor $\Delta$. It
vanishes at $r=0$, that is so because at the center the effective radial and transverse pressures coincide. On the other hand, as the radius increases the values of these quantities drift apart, and therefore the anisotropy increases toward the surface of the object. Furthermore Fig.
(\ref{zeldovich}) shows that Zeldovich's condition is satisfies everywhere inside the object.
Table \ref{table2} reports values corresponding to central and surface density which are according to the expected ranges for a star formed by a quark fluid, also radial pressure and Zeldovich's condition are reported at the center of the star.  

\subsection{Energy conditions}
\begin{figure}[H]
\includegraphics[scale=2.0]{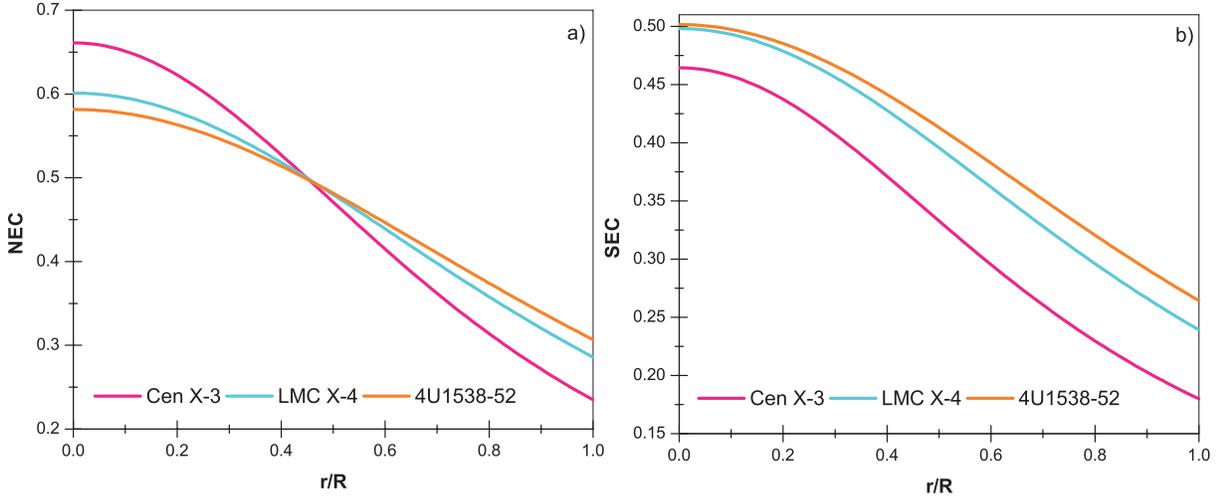}
\caption{Null and Strong energy  conditions against the dimensionless radial coordinate
for Cen X-3, LMC X-4 and 4U1538-52 for
the parameters given in table \ref{table1}.}
\label{nec}
\end{figure}

\begin{figure}[H]
\includegraphics[scale=2.0]{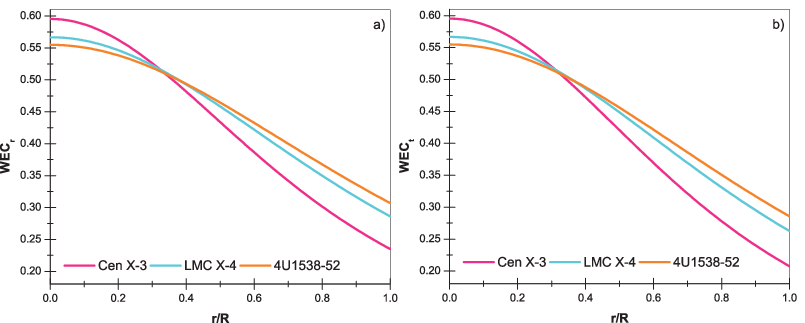}
\caption{Weak energy conditions against the dimensionless radial coordinate
for Cen X-3, LMC X-4 and 4U1538-52 for
the parameters given in table \ref{table1}.}
\label{wec}
\end{figure}

\begin{figure}[H]
\includegraphics[scale=2.0]{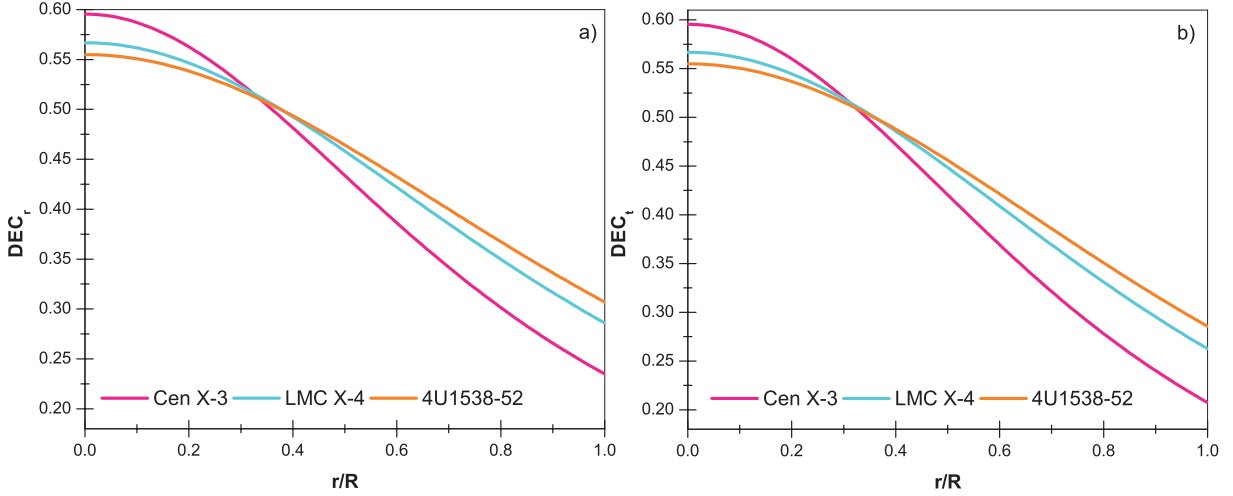}
\caption{Dominant energy conditions against the dimensionless radial coordinate
for Cen X-3, LMC X-4 and 4U1538-52 for
the parameters given in table \ref{table1}.}
\label{dec}
\end{figure}

Within the anisotropic matter distribution the energy should be positive. In order to ensure it, the energy-momentum tensor has to obey the null energy condition (NEC), the weak energy condition (WEC) in both tangential and radial direction, the strong energy condition (SEC) and the dominant energy conditions (DEC) in both tangential and radial direction\cite{ponceLeon1,visserbook}:
\begin{enumerate}
    \item (NEC): $\rho \geq 0$.
    \item (WEC): $\rho-p_{t}\geq 0$, $\rho-p_{r}\geq 0$ .
    \item (SEC): $\rho-2p_{t}-p_{r}\geq 0$.
    \item (DEC): $\rho -|p_{r}|\geq0$, $\rho -|p_{t}|\geq0$.
\end{enumerate}
Figures (\ref{nec}), (\ref{wec}) and (\ref{dec}) shown that all the above inequalities are satisfied within the object. Therefore we have a well behaved energy-momentum tensor.

\begin{figure}[H]
\centering
\includegraphics[scale=1.2]{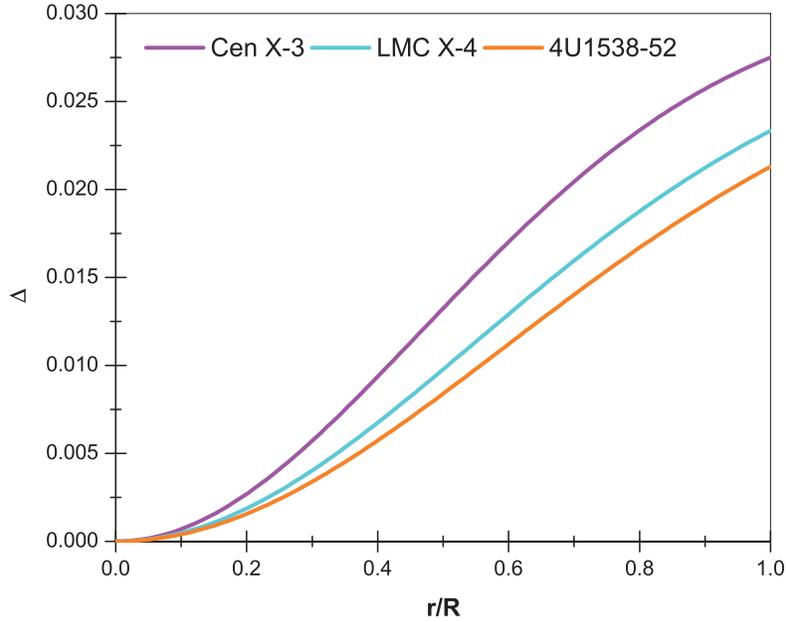}
\caption{The anisotropy factor $\Delta$ against the dimensionless radial coordinate
for Cen X-3, LMC X-4 and 4U1538-52 for
the parameters given in table \ref{table1}.}
\label{anisotropy}
\end{figure}

\begin{table}[H]
\caption{Constant parameters calculated for radii and mass for some strange star candidates.}
\label{table1}
\begin{tabular*}{\textwidth}{@{\extracolsep{\fill}}lrrrrrl@{}}
\hline
Strange star & \multicolumn{1}{c}{radii $(R)/$} & \multicolumn{1}{c}{$M/$} & \multicolumn{1}{c}{$B/$} & \multicolumn{1}{c}{$C/$}&
\multicolumn{1}{c}{$A$ (dimen-}
 \\
candidates &$(km)$& $M_{\odot}$&$(\times10^{-3}km^{-2})$&$(km)$& sionless) \\
\hline
Cen X$-$3 & 9.178 & 1.49 & 1.96552 & 13.2704& -0.55610 \\
\hline
LMC X$-$4 &8.301 & 1.04 & 1.68619 & 13.6626& -0.67807 \\ \hline
4U 1538-52 & 7.866 & 0.87 & 1.60368 & 13.7792 & -0.72182 \\
\hline
\end{tabular*}
\end{table}

\begin{table}[H]
\caption{Some physical parameters calculated for radii and mass for some strange star candidates.}
\label{table2}
\begin{tabular*}{\textwidth}{@{\extracolsep{\fill}}lrrrrrl@{}}
\hline
Strange star & \multicolumn{1}{c}{$\rho(0)/$} & \multicolumn{1}{c}{$\rho(R)/$} & \multicolumn{1}{c}{$p_{r}(0)/$} & \multicolumn{1}{c}{$p_{r}(0)/$}
 \\
candidates &$(\times 10^{15} gcm^{-3})$& $(\times 10^{15} gcm^{-3})$&$\rho(0)$&$(\times 10^{35} dyne/ cm^{2})$\\
\hline
Cen X$-$3 & 1.75323 & 0.62295 & 0.09914 & 1.56427 \\
\hline
LMC X$-$4 &1.36773 &0.65050 &0.05712 &0.70309\\
\hline
4U 1538-52 & 1.25837 & 0.66400 & 0.04582 & 0.51895 \\ 
\hline
\end{tabular*}
\end{table}

\subsection{Mass function and compactness factor}
In the study of spherically symmetric and static distributions associated with a perfect fluid, the maximum limit of the mass-radius ratio must satisfy the following upper bound $u=\tilde{M}/R<4/9$ (in the units $c=G=1$) \cite{buchdahl}. Nevertheless, in the presence of an anisotropic matter distribution this limit is more general \cite{andreasson}. However, it can be obtained from the relationship between $e^{\lambda}$ and mass function
$m(r)$, i.e.
\begin{equation}\label{massfunction}
e^{-\lambda}=1-\frac{2m(r)}{r}.    
\end{equation}
Thus, we get the expression of the mass function as
\begin{equation}\label{mass}
m(r)=\frac{8B^{2}C^{2}r^{3}}{1+16B^{2}C^{2}r^{2}}  
\end{equation}
The compactness factor $u$ for the model is obtained as
\begin{equation}\label{compactness}
u(r)=\frac{2m(r)}{r}=\frac{16B^{2}C^{2}r^{2}}{1+16B^{2}C^{2}r^{2}} .    
\end{equation}
Fig. (\ref{mass}) displays the mass function and the compactness factor $u$. Table \ref{table3} shows the corresponding compactness factor for each strange star candidate chosen in this work. 

\subsection{Surface and central redshift}

\begin{figure}[H]
\centering
\includegraphics[scale=2.0]{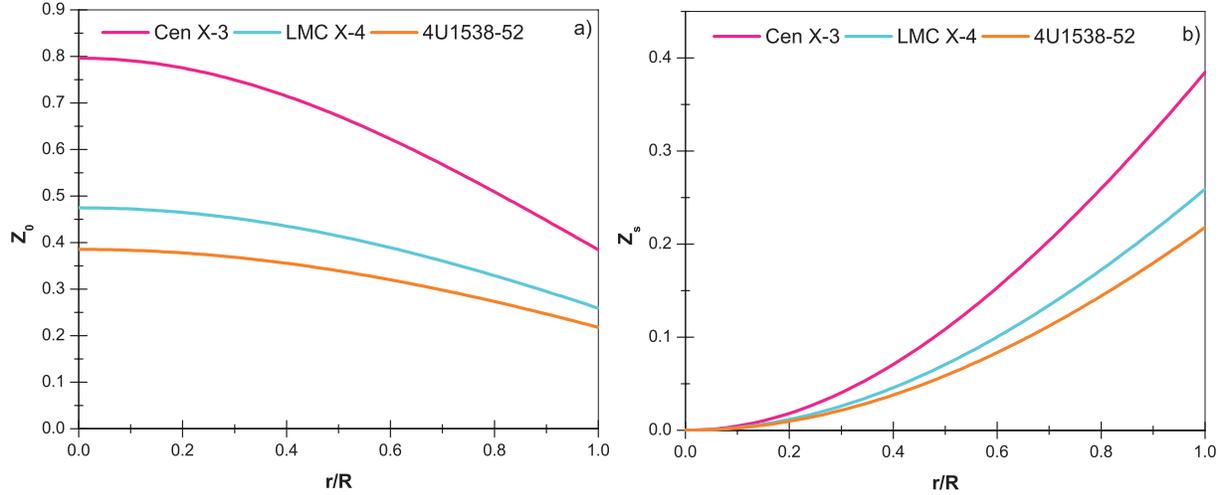}
\caption{Surface $Z_{s}$ and central redshift $Z_{0}$ against the dimensionless radial coordinate
for Cen X-3, LMC X-4 and 4U1538-52 for
the parameters given in table \ref{table1}.}
\label{redshift}
\end{figure}

Related to the surface redshift, it can be calculated using the compactness factor $u$ given by (\ref{compactness}), as follows
\begin{equation}
Z_{s}=e^{\lambda(r)/2}-1=\frac{1-\sqrt{1-2u}}{\sqrt{1-2u}},    
\end{equation}
explicitly it reads
\begin{equation}
Z_{s}=\sqrt{\left(1+16B^{2}C^{2}r^{2}\right)}-1.    
\end{equation}
The presence of a positive anisotropy factor $\Delta>0$ does not impose an upper limit on the surface redshift $Z_{s}$. In distinction with what happens in the case of isotropic distributions, where the maximum value that the surface redshift $Z_{s}$ can reaches is $Z_ {s} = 4.77$ \cite{bowers}. Therefore, the surface redshift for anisotropic matter configurations is greater than its isotropic counterpart. On the other hand the central redshift $Z_{0}$ can be obtained as follows
\begin{equation}
Z_{0}=e^{-\nu(r)/2}-1=\frac{1}{\sqrt{\left(A-Br^{2}\right)^{2}}}-1.    
\end{equation}
Figure (\ref{redshift}) shows that both the surface $Z_{s}$ and the central $Z_{0}$ redshift are positive and bounded within the star. We report in table \ref{table3} the corresponding values for the central and the surfce redshift.



\begin{figure}
\centering
\includegraphics[scale=2.0]{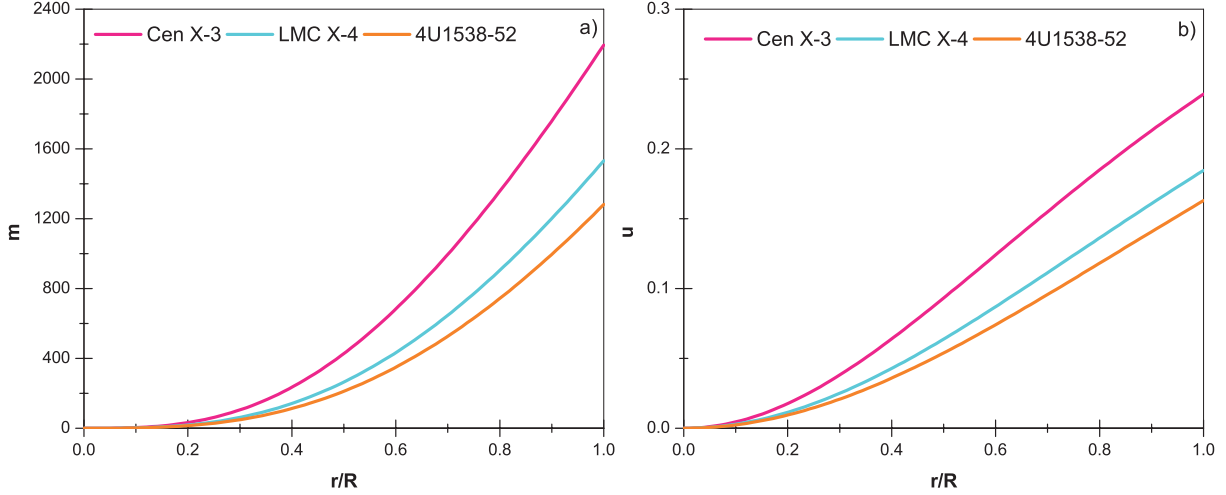}
\caption{Mass function $m(r)$ and compactness factor $u$ against the dimensionless radial coordinate
for Cen X-3, LMC X-4 and 4U1538-52 for
the parameters given in table \ref{table1}.}
\label{mass}
\end{figure}

\section{Equilibrium and stability conditions}\label{section6}

\subsection{Causality condition and Abreu's criterion}

An admissible anisotropic solution to the Einstein's field equations must satisfies causality condition \i.e both the radial $v_{r}$ and tangential $v_{t}$ sound speeds inside the object are less than the speed of light $c$ (in relativistic geometrized units the speed of light becomes $c=1$ ). Explicitly it reads 
\begin{eqnarray}
v_{r}(r)&=&\sqrt{\frac{dp_{r}(r)}{d\rho(r)}}\leq1 \\
v_{t}(r)&=&\sqrt{\frac{dp_{t}(r)}{d\rho(r)}}\leq1.
\end{eqnarray}
As shown in figure \ref{velocities} both speeds fulfill the above requirement.

The stability of anisotropic stars under the radial
perturbations is studied by using the concept known
as cracking \cite{herrera}. Based on it Abreu et. al. \cite{abreu} established another alternative to study the stability of a self-gravitating anisotropic fluid sphere. This approach states that the region is potentially stable where the radial speed $v_{r}$ of sound is greater than the 
transverse $v_{t}$ speed of sound. This implies that there is no change in sign $v^{2}_{t}-v^{2}_{r}$. The later assumption is equivalent to $0\leq|v^{2}_{t}-v^{2}_{r}|<1$.\\
We note from figure \ref{velocities} that radial speed of sound is always greater than
transverse speed of sound and also from figure \ref{abreu} panel $a)$ shows that there is no change in sign. On the other hand, panel $b)$ shows $0\leq|v^{2}_{t}-v^{2}_{r}|<1$ everywhere inside the star.
These features represent that the proposed physical model is stable.

\subsection{Relativistic adiabatic index}
\begin{figure}[H]
\centering
\includegraphics[scale=1.2]{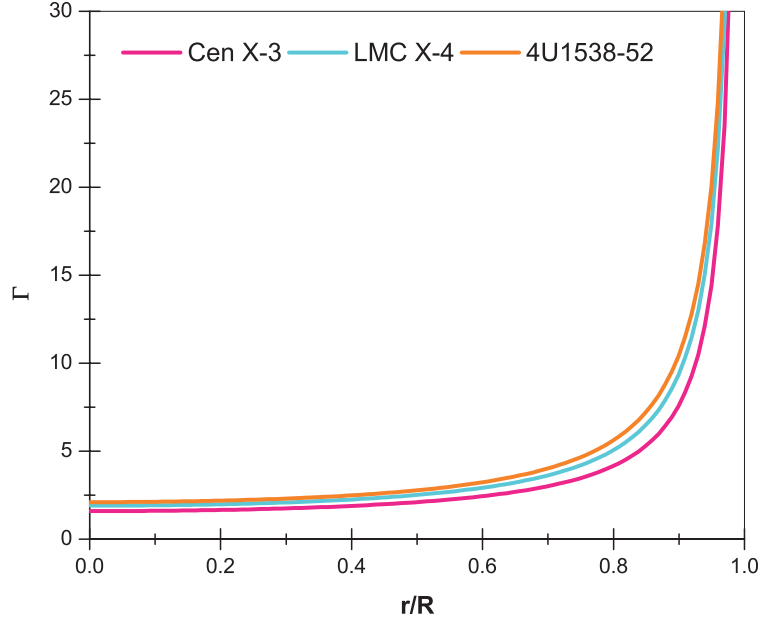}
\caption{Adiabatic index $\Gamma$ against the dimensionless radial coordinate
for Cen X-3, LMC X-4 and 4U1538-52 for
the parameters given in table \ref{table1}.}
\label{adiabatic}
\end{figure}

Another relevant aspect in the study of stability of compact anisotropic matter distriburions is related with the relativistic adiabatic index $\Gamma$ inside the compact object \cite{bondi,heintz}.\\
It is well known from the studies about Newtonian isotropic fluid spheres that the collapsing condition correspond to $\Gamma<4/3$. In distinction with the relativistic counterpart this condition becomes \cite{chan1,chan2}
\begin{equation}\label{adibatic}
\Gamma<\frac{4}{3}+\left[\frac{1}{3}\kappa\frac{\rho_{0}p_{r0}}{|p^{\prime}_{r0}|}r+\frac{4}{3}\frac{\left(p_{t0}-p_{r0}\right)}{|p^{\prime}_{r0}|r}\right]_{max},    
\end{equation}
where $\rho_{0}$, $p_{r0}$ and $p_{t0}$ are the initial density, radial and tangential pressure when the fluid is in static equilibrium. The second term in the right hand side represents the relativistic corrections to the Newtonian perfect fluid and the third term is the contribution due to anisotropy. It is clear from (\ref{adibatic}) that in the case of a non-relativistic matter distribution and taking $p_{r}$ be equal to $p_{t}$ \i.e $\Delta=0$, the bracket vanishes and we recast the collapsing Newtonian limit $\Gamma<4/3$. On the other hand, Heintzmann and Hillebrandt \cite{heintz} showed that in the presence of a positive and increasing anisotropy factor $\Delta=p_{t}-p_{r}>0$, the stability condition for a relativistic compact object is given by $\Gamma>4/3$, this is so because positive anisotropy factor
may slow down the growth of instability. Due to the fact that gravitational collapse occurs in the radial direction, it is enough to analyze the adiabatic index in such direction. So, the adiabatic index is given by \cite{chan3}  
\begin{equation}
\Gamma=\frac{\rho+p_{r}}{p_{r}}\frac{dp_{r}}{d\rho}.    
\end{equation} 
We can see from figure \ref{adiabatic} that the model is in complete agreement with the condition $\Gamma>4/3$. Then the model is stable. Table \ref{table3} shows the exact values for $\Gamma$, all of them being greater than $4/3$ everywhere inside the configuration.
\subsection{Equilibrium under three different forces}

The equilibrium of the system lies on the Tolman-Oppenheimer-Volkoff (TOV) equation \cite{tolman,oppenheimer}. In this case the equilibrium of the anisotropic fluid sphere is under three different forces. This forces are the hydrostatic force $F_{h}$, the gravitational force $F_{g}$ and the anisotropic repulsive force $F_{a}$ introduced by the presence of a positive anisotropy factor $\Delta$. In fact, the hydrostatic force $F_{h}$ and the anisotropic repulsive force $F_{a}$ counterbalance the gravitational force $F_{g}$. Therefore, the collapse of the compact object to a point singularity may be avoided during the gravitational collapse. In conclusion, the presence of anisotropies within the stellar configuration enhance the equilibrium of the system. \\
Then the TOV equation describing the equilibrium condition for an anisotropic 
fluid distribution is given by

\begin{equation}\label{TOV}
-\underbrace{\frac{\nu^{\prime}}{2}\left(\rho+p_{r}\right)}_{F_{g}} - \underbrace{\frac{dp_{r}}{dr}}_{F_{h}}+\underbrace{\frac{2}{r}\Delta}_{F_{a}}=0.
\end{equation}
The explicit expressions of these forces are
\begin{equation}
F_{a}=\frac{8B^3C^2r\left(8B^2C^2r^2-8ABC^2-1\right)}{\pi\left(Br^2-A\right)\left(16B^2C^2r^2+1\right)^2},    
\end{equation}
\begin{equation}
F_{g}=\frac{B^2r\left(8ABC^{2}-24B^{2}C^{2}r^{2}-1\right)}{\pi\left(Br^2-A\right)\left(16B^2C^2r^2+1\right)^2},    
\end{equation}
\begin{equation}
\begin{split}
F_{h}=\frac{-B^{2}r}{\pi\left(Br^2-A\right)\left(16B^2C^2r^2+1\right)^2}\bigg[64B^4C^4r^4-128 & \\ \times AB^3C^4r^2 +64A^2B^2C^4-32B^2C^2r^2+16ABC^2-1\bigg].  
\end{split}
\end{equation}

\begin{figure}[H]
\centering
\includegraphics[scale=2.0]{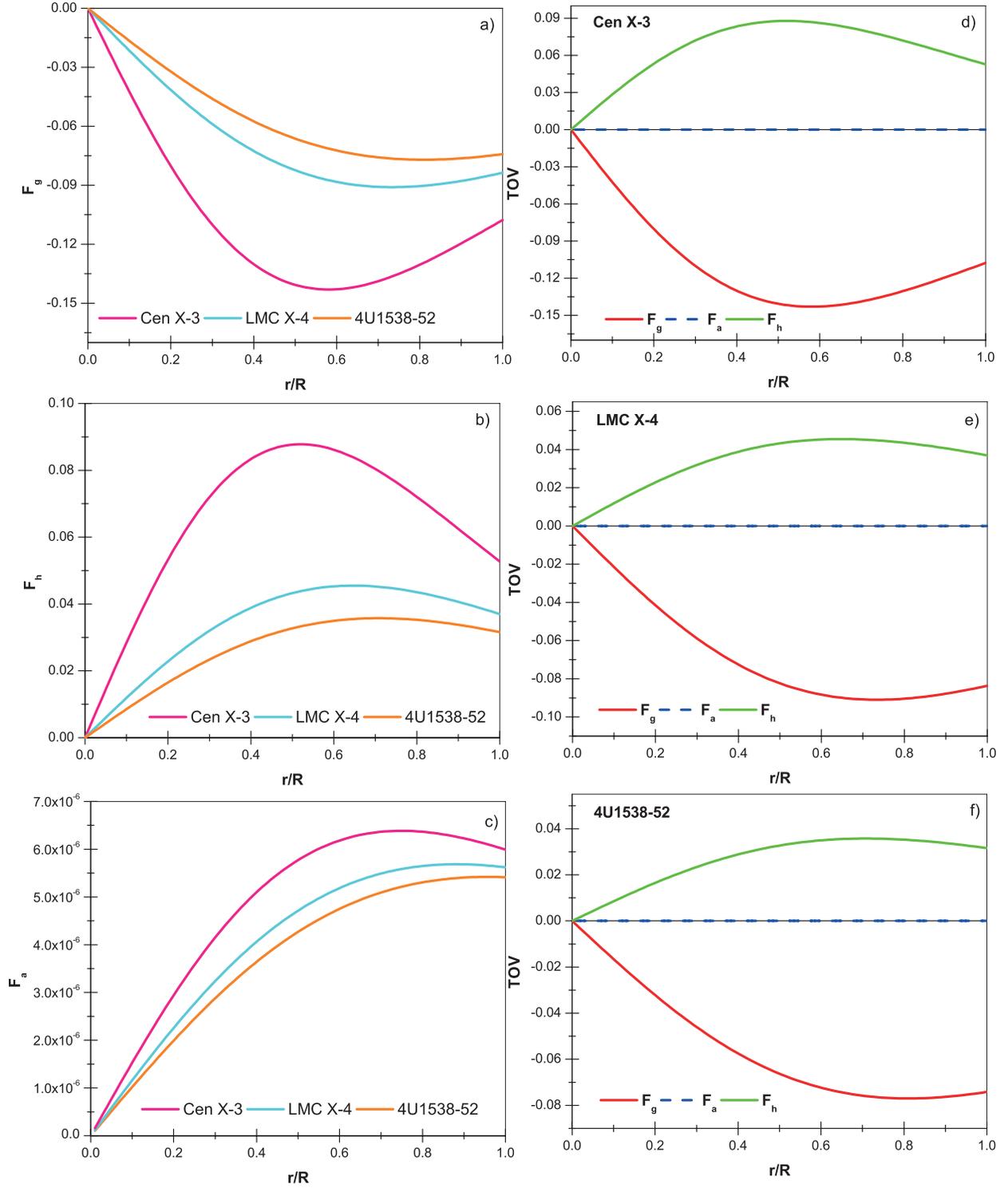}
\caption{TOV equation against the dimensionless radial coordinate
for Cen X-3, LMC X-4 and 4U1538-52 for
the parameters given in table \ref{table1}.}
\label{tov}
\end{figure}

We can observe from figure (\ref{tov}) that the model is in equilibrium under the mentioned forces.

\begin{figure}[H]
\includegraphics[scale=2.0]{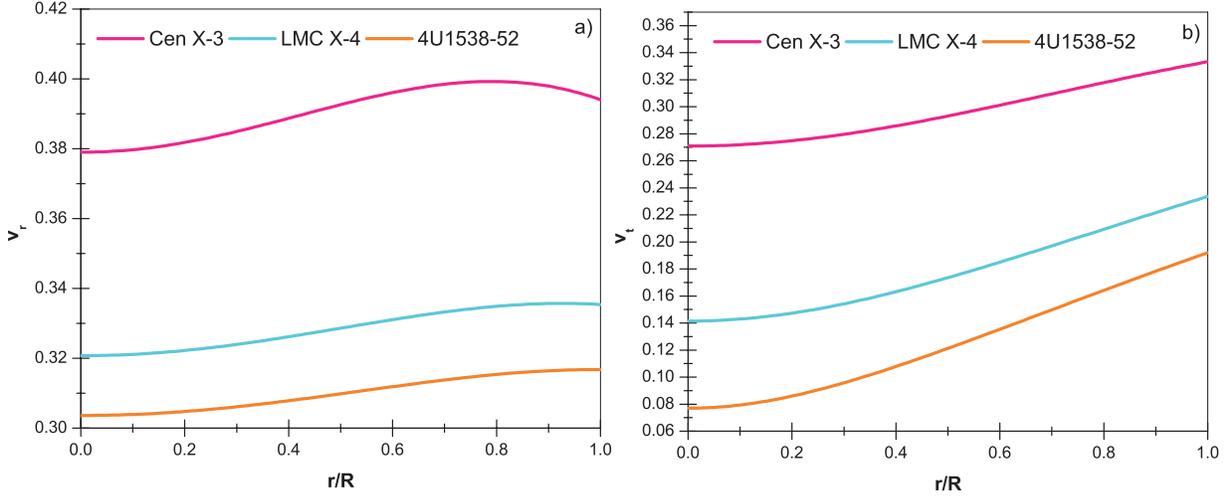}
\caption{Radial $v_{r}$ panel $a)$ and tangential $v_{t}$ panel $b)$ sound speeds against the dimensionless radial coordinate
for Cen X-3, LMC X-4 and 4U1538-52 for
the parameters given in table \ref{table1}.}
\label{velocities}
\end{figure}

\begin{figure}[H]
\includegraphics[scale=2.0]{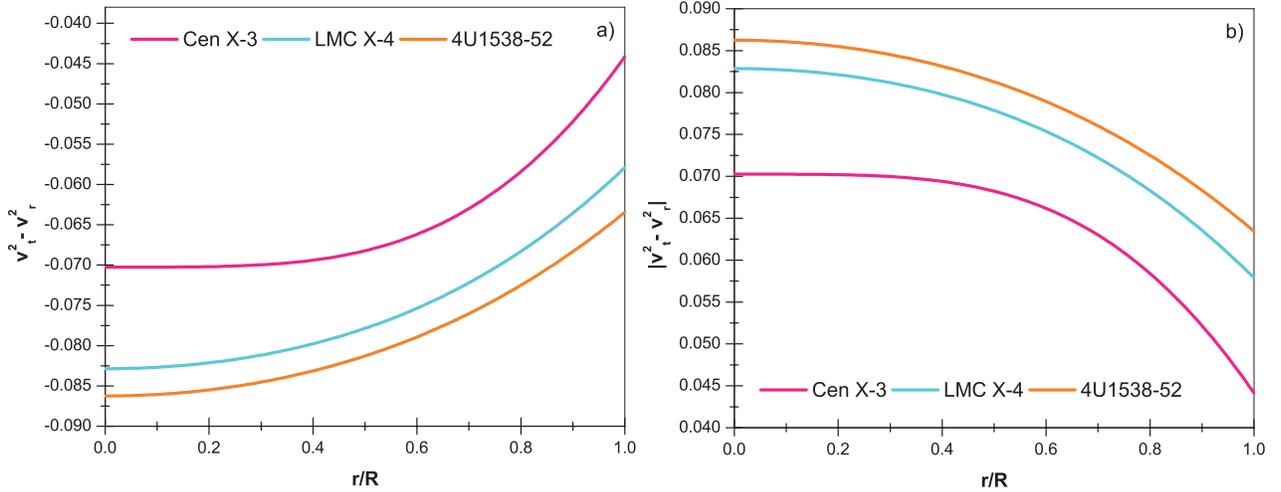}
\caption{Variation of the stability factor against the dimensionless radial coordinate
for Cen X-3, LMC X-4 and 4U1538-52 for
the parameters given in table \ref{table1}.}
\label{abreu}
\end{figure}

\begin{table}[H]
\caption{Some physical parameters calculated for radii and mass for some strange star candidates.}
\label{table3}
\begin{tabular*}{\textwidth}{@{\extracolsep{\fill}}lrrrrrl@{}}
\hline
Strange star & \multicolumn{1}{c}{$u$} & \multicolumn{1}{c}{$Z_{0}$ } & \multicolumn{1}{c}{$Z_{s}$} & \multicolumn{1}{c}{$\Gamma$}&
 \\
candidates&(compactness factor)& (central redshift)& (surface redshift)&(adibatic index)\\
\hline
Cen X$-$3 & 0.239 & 0.79630 & 0.38453 & 1.59261 \\
\hline
LMC X$-$4 &0.185 &0.47476 &0.25903 &1.90396\\
\hline
4U 1538-52 & 0.163 & 0.38538 & 0.21795 & 2.10398 \\ 
\hline
\end{tabular*}
\end{table}

\section{Concluding remarks}\label{section7}
Embedding class one spacetime using Karmarkar's condition has been proved  as a simple and powerful tool to obtain solution of the Einstein's field equations. Despite its simplicity the obtained model describing anisotropic fluid spheres, fulfills all the requirements in order to be an admissible solution from both the physical and mathematical point of view. It is remarkable to mention that the present model is free from physical and geometric singularities (Fig. \ref{pots}), furthermore the energy density $\rho$, the radial $p_{r}$ and tangential $p_{t}$ pressures are completely finite and positive quantities within the stellar configuration (Fig. \ref{thermo}). Moreover, the energy-momentum tensor associated with this anisotropic matter distribution is well behaved, it can be seen from the graphical analysis of the energy conditions (Fig. \ref{nec}, \ref{wec} and \ref{dec}), additionally the central $Z_{0}$ and the surface $Z_{s}$ redshift are both positive and bounded (Fig. \ref{redshift}).  

On the other hand, the system is in equilibrium under three different forces, namely the hydrostatic force $F_{h}$, the gravitational force $F_{g}$ and the anisotropic force $F_{a}$ (Fig. \ref{tov}). The latest one causes a repulsive force that counteracts the gravitational gradient, this is so because we are in presence of a positive anisotropy factor $\Delta$ as can be seen in Fig \ref{anisotropy}. As was pointed out by Ruderman \cite{ruderman} and Canuto \cite{canuto} in their early theoretical works, anisotropy can arise in ultra high density ranges. Table \ref{table1} shows that the central energy density is within this range, which it in complete agreement with many other reported results in the literature \cite{schunk,ray,usov,mello,negreiros,varela,rahaman1,rahaman2,rahaman3,rahaman,hassan1,maurya7,bhar1,rahaman4,kileba1,hassan,shee,monadi,smaurya1,smaurya2,deb1,Ovalle8,Ovalle,role,jasim,Tello1,Gabbanelli,camilo,sharif,Ovalle9,Tello,Tello2,deb,kileba}. Respect to the stability of the model, it was studied analyzing Abreu's criterion and the relativistic adiabatic index $\Gamma$. From the point of view of Abreu's criterion the model is completely stable, because the radial sound speed $v_{r}$ is always greater than the transverse sound speed $v_{t}$ everywhere within the star (Fig. \ref{velocities}), besides there is no change in sign  $v^{2}_{t}-v^{2}_{r}$ and stability factor $|v^{2}_{t}-v^{2}_{r}|$ lies between 0 and 1(Fig. \ref{abreu}). Finally, the relativistic adibatic index $\Gamma$ is greater than $4/3$ and increases monotonically outward, it means that the system is stable.

\section{Acknowledgements}
F. Tello-Ortiz thanks the financial support of the the project ANT-1756 at the Universidad de Antofagasta, Chile.

\end{document}